\documentclass[aps,prl,reprint,floatfix]{revtex4-1}

\usepackage{amssymb}
\usepackage{amsmath}
\usepackage{amsfonts}
\usepackage{graphicx}

\newcommand{\fancy}{\mathcal}

\DeclareMathOperator*{\argmin}{arg\,min}
\DeclareMathOperator*{\argmax}{arg\,max}

\newcommand{\FE}{\kappa}

\newcommand{\lbrs}{\left[}
\newcommand{\rbrs}{\right]}

\renewcommand{\vec}[1]{\boldsymbol{#1}}

\newcommand{\beq}{\begin{eqnarray}}
\newcommand{\eeq}{\end{eqnarray}}

\newcommand{\tr}{\text{Tr}}
\newcommand{\Tr}[1]{{{\text{Tr}}\lbrs #1 \rbrs}}

\newcommand{\half}{\frac{1}{2}}

\newcommand{\Hx}{\text{Hx}}

\newcommand{\crm}{\text{c}}
\newcommand{\Hrm}{\text{H}}

\newcommand{\Hxc}{\text{Hxc}}

\newcommand{\W}{{\cal W}}

\newcommand{\Ts}{{\cal T}_{s}}

\newcommand{\E}{{\cal E}}
\newcommand{\F}{{\cal F}}

\newcommand{\EHxc}{\E_{\Hxc}}
\newcommand{\EHx}{\E_{\Hx}}
\newcommand{\Ec}{\E_{\crm}}

\newcommand{\SD}{{\text{SD}}}
\newcommand{\DD}{{\text{DD}}}

\newcommand{\DDA}{{\text{DDA}}}

\newcommand{\gs}{\text{gs}}
\newcommand{\ts}{\text{ts}}
\renewcommand{\ss}{\text{ss}}

\newcommand{\pr}{^{\prime}}

\renewcommand{\vr}{\vec{r}}

\newcommand{\vrp}{\vec{r}\pr}

\newcommand{\iket}[1]{|#1\rangle}

\newcommand{\ibraketop}[3]{\langle#1|#2|#3\rangle}
\newcommand{\ibkouter}[1]{|#1\rangle\langle#1|}

\newcommand{\iout}{\ibkouter}

\newcommand{\EXX}{{\text{EXX}}}

\newcommand{\nh}{\hat{n}}
\newcommand{\Th}{\hat{T}}
\renewcommand{\th}{\hat{t}}
\newcommand{\Wh}{\hat{W}}
\newcommand{\vh}{\hat{v}}

\newcommand{\Hh}{\hat{H}}
\newcommand{\hh}{\hat{h}}

\newcommand{\Gammah}{\hat{\Gamma}}

\newcommand{\vsE}{v_s^\FE}

\newcommand{\Tb}{\bar{T}}
\newcommand{\Eb}{\bar{E}}
\newcommand{\FEXXb}{\bar{F}^{\EXX}_{\FE}}
\newcommand{\FEXX}{{F}^{\EXX}_{\FE}}

\newcommand{\Lambdab}{\bar{\Lambda}}

\newcommand{\nt}{\tilde{n}}

\usepackage[normalem]{ulem}
\usepackage{color}
\usepackage{xcolor}

\definecolor{Mygrey}{gray}{0.80}

\begin{document}
\title{Density-driven correlations in many-electron ensembles: \\ theory and application for excited states}
\author{Tim Gould}\affiliation{Qld Micro- and Nanotechnology Centre, %
  Griffith University, Nathan, Qld 4111, Australia}
\author{Stefano Pittalis}\affiliation{CNR-Istituto Nanoscienze, Via
  Campi 213A, I-41125 Modena, Italy}

\begin{abstract}
  Density functional theory can be extended to excited states by means
  of a unified variational approach for passive state ensembles. This
  extension overcomes the restriction of the typical density
  functional approach to ground states, and offers useful formal and
  demonstrated practical benefits. The correlation energy
  functional in the generalized case acquires higher complexity
  than its ground state counterpart, however. Little is known
  about its internal structure nor how to effectively approximate
  it in general. Here we show that such a functional
  can be broken down into natural components, including what we call
  ``state-'' and ``density-driven'' correlations, with the
  former amenable to conventional approximations, and the
  latter being a unique feature of ensembles. Such a decomposition,
  summarised in eq.~\eqref{eqn:EcE}, provides us with a pathway to
  general approximations that are able to routinely handle low-lying
  excited states.
  The importance of density-driven correlations is demonstrated,
  an approximation for them is introduced and shown to be useful.
\end{abstract}

\maketitle

Electronic structure theory has transformed the study of chemistry,
materials science and condensed matter physics, by enabling
quantitative predictions using computers.  But a general solution to
the many-electron problem remains elusive, because the
electron-electron interactions imply highly non-trivial correlations
among the relevant degrees of freedoms.  Out of the numerous
electronic structure methodologies, density functional theory
\cite{HohenbergKohn,KohnSham,Jones2015-Perspective} (DFT) has become
the dominant approach
thanks to its balance between accuracy and speed, achieved by using
the electron density as the basic variable, then mapping
the original interacting problem onto an auxiliary non-interacting
problem.

DFT gives access to ground states, but not excited states,
meaning alternatives must be used for important processes like
photochemistry or exciton physics \cite{Matsika2018-ChemRev}.
Its time-dependent extension (TDDFT) does offer
access to excited states at reasonable cost
\cite{RungeGross,Casida2012-Review}, and
is thus commonly employed for this purpose. Routine applications
of TDDFT reuse ground-state approximations by evaluating them on the
instantaneous density, the so-called adiabatic approximation.
This approach fails badly, however, when many-body correlations defy
a time-dependent mean-field picture, including for important charge
transfer excitations \cite{Ullrich2006,Maitra2017-CT}.

One highly promising alternative involves tackling both ground
and excited eigenstates by means of one and the same density
functional approach \cite{Theophilou79,Gross1988-1,Gross1988-2,Oliveira1988-3},
using ensemble DFT (EDFT). EDFT is appealing because it can
automatically deal with otherwise difficult orthogonality conditions
and can potentially tap into more than 30 years of density
functional approximation development.
EDFT has been shown to solve problems that are difficult for TDDFT,
such as charge transfers, double excitations, and conical
intersections
\cite{Filatov1999-REKS,Filatov2015-Double,Filatov2015-Review,Filatov2016,%
Franck2014,Deur2017,%
Pribram-Jones2014,Yang2014,Yang2017-EDFT,%
Gould2018-CT,Sagredo2018}.

Consolidating the preliminary success of EDFT into useful
approximations requires further understanding of how many-body
correlations get encoded in EDFT and how they can be approximated
generally. The correlation energy of many-electron ground states
is traditionally divided into dynamical (weak) and static (strong)
correlations. This decomposition is by no means unambiguous, yet is
very useful both for designing, and understanding the limitations
of, approximations \cite{Ghosh2018}.
Both static and dynamic correlations are also present in ensembles.
But the internal structure of the correlation energy functional for
ensembles is, by necessity, more complex. Little is known about its
specific properties and quirks.

In this Letter, we {reveal} a decomposition of
the ensemble correlation energy that lends itself both to an exact
evaluation {\em and} to a universal approximation scheme. {Our
  decomposition uncovers components of the correlation energy in
  multi-state ensembles, that will be missed by direct reuse of
  existing density functional approximations on pure-state
  contributions. We show that the additional components are unique
  features of EDFT and can lead to significant errors, if ignored. We
  thus point out a {crucial missing step on the path} to
  upgrade existing approximations for correlations.}

The  components revealed through our decomposition --
{\em density-driven correlations} -- have so far gone unnoticed,
and are similar to, but not the same as density-driven
errors of approximations \cite{Kim2013-DensDriven}.
Ultimately, these components appear because the Kohn-Sham scheme in
EDFT provides the exact overall ensemble particle density,
but not the density of each state in the ensemble.
Our approach makes use of recent results on the Hartree-exchange
component of the ensemble energy \cite{Gould2017-Limits} and
introduces a generalization of the Kohn-Sham machinery. We shall describe our construction first formally and
then also by means of direct applications. {The relevance of the density-driven
correlation is thus established unambiguously for
prototypical cases}.

{\em A primer on EDFT:} 
For a given electron-electron interaction
strength $\lambda$, external potential $v$, and set of weights $\W$
one can find\cite{Gross1988-1} an ensemble density matrix,
\begin{align}
  \Gammah^{\lambda}[v;\W]=\sum w_{\FE}\iout{\FE^{\lambda}}
  \equiv\argmin_{\Gammah\to\W}\Tr{\Gammah\Hh^{\lambda}[v]},
  \label{eqn:GammanW}
\end{align}
so that $\fancy{E}^{\lambda}[v;\W]=\tr[\Gammah^{\lambda}\Hh^{\lambda}[v]]
=\sum_{\FE}w_{\FE}E_{\FE}^{\lambda}$ is the energy of the ensemble system.
Here $\W=\{w_{\FE}\}$ describes a set of non-negative weights
that obey $\sum_{\FE} w_{\FE} =1$. A consequence of
\eqref{eqn:GammanW} is that $\iket{\FE^{\lambda}}$ are eigenfunctions of
$\Hh^{\lambda}[v]=\Th + \lambda\Wh + \int \nh(\vr)v(\vr) d\vr$
sorted so that $w_{\FE}\leq w_{\FE'}$ for
eigenvalues $E^{\lambda}_{\FE}>E^{\lambda}_{\FE'}$ 
where $E^{\lambda}_{\FE}=\ibraketop{\FE}{\Hh}{\FE}$, making
the ensemble a passive state from which no work can be
extracted\cite{Perarnau-Llobet2015}.
We can, without loss of generality, assign
equal weights whenever interacting states are degenerate.
{Excitation energies can be found via derivatives
or differences of $\fancy{E}^1$
with respect to relevant excited state weights $w_{\FE>0}$
\cite{Theophilou79,Gross1988-2,Gould2018-CT,Deur2019}.}

By the Gross-Oliveira-Kohn (GOK)
theorems \cite{Gross1988-1,Gross1988-2,Oliveira1988-3} and
the usual assumption that all densities of interest are ensemble
$v$-representable, there exists a potential,
$v^{\lambda}[n;\W] \equiv \argmax_u\{
\fancy{E}^{\lambda}[u;\W] - \int n u d\vr \},$
that is a unique functional of $n$ and $\W$.
Notice here we allow $\lambda$ to vary while keeping $n$ constant to
connect ``adiabatically'' the non-interacting ($\lambda = 0$,
$v^0\equiv v_s$) with the fully interacting limits
($\lambda =1$, $v^1\equiv v$).
To simplify discussion, we further restrict to the ``strong
adiabatic'' case that the ordering of occupied states ($w_{\FE}>0$)
as $\lambda\to 0^+$ is the same as at $\lambda=1$, i.e. that the
energy ordering of low-lying states is adiabatically preserved. This
is true in the cases considered here and the majority of cases
amenable to EDFT -- exceptions, we suspect, may include magnetic
states such as those with relevant orbital degeneracies in
combination with strong and spin-orbit interactions. Our consequent
discussion should be extended to cover such exceptions.

Since $v^{\lambda}\to n$ and $n\to v^{\lambda}$ are unique mappings
at all relevant $\lambda$, for weights $\W$,
we can define the universal ensemble density functional
\begin{align}
  \F^{\lambda}[n]\equiv&\sum_{\FE}w_{\FE}
  \ibraketop{\FE^{\lambda}}{\Th+\lambda\Wh}{\FE^{\lambda}}
  \equiv\tr[\Gammah^{\lambda}(\Th+\lambda\Wh)]
  \label{eqn:Flambda}
\end{align}
where $\iket{\FE^{\lambda}}$ are eigenstates of $[\Th+\lambda \Wh +
  \vh^{\lambda}]\iket{\FE^{\lambda}}
=E_{\FE}^{\lambda}\iket{\FE^{\lambda}}$, $\Gammah^{\lambda}=\sum
w_{\FE}\iout{\FE^{\lambda}}$ and $\tr[\Gammah^{\lambda} \nh] =
\sum w_{\FE}\ibraketop{\FE^{\lambda}}{\nh}{\FE^{\lambda}} =n$. 
For brevity, we now drop explicit references to $\W$.

Making use of the Kohn-Sham (KS) ensemble, the interacting universal
functional at $\lambda=1$ ($ \F[n] \equiv \F^{1}[n] $) can be
decomposed as
$
  \F[n] =  \Ts[n] +  \EHx[n] +  \Ec[n]
$
where $\Ts[n]$, $\EHx[n]$ and $\Ec[n]$ are the ensemble KS kinetic,
Hartree-exchange (Hx) energy, and correlation energy functionals.
We shall focus on cases involving degeneracies for different spin
states but no ambiguities for the spatial degree-of-freedom --
this is sufficient for elucidating the main points of this work. Thus,
the KS kinetic and Hx energy are given, respectively, by
\begin{align}
  \Ts[n] \equiv& \F^0[n]=\sum w_{\FE}T_{s,\FE}[n],
  \label{eqn:Ts}
  \\
  \EHx[n]\equiv& \lim_{\lambda\to 0^+}\frac{\F^{\lambda}[n]-\F^0[n]}{\lambda}
  =\sum_{\FE}w_{\FE}\Lambda_{\Hx,\FE}[n],
  \label{eqn:EHx}
\end{align}
where $T_{s,\FE}=\ibraketop{\FE^{0+}}{\Th}{\FE^{0+}}$,
$\Lambda_{\Hx,\FE}=\ibraketop{\FE^{0+}}{\Wh}{\FE^{0+}}$.
$\iket{\FE^{0+}}$ are orthogonal (formally non-interacting) eigenstates  
as well as proper spin eigenstates -- they thus may be
linear combinations of Slater determinants which
``optimize'' $\EHx$ \cite{Gould2017-Limits}.
Of relevance to our discussion are the following three facts:
(1) $\Ts$ and $\EHx$ are functionals of a shared
set of occupied one-body orbitals $\phi_i[n](\vr)$ obeying
$[\th + v_s[n]] \phi_i[n](\vr)=\epsilon_i[n]\phi_i[n](\vr)$;
(2) Some states
(e.g. singlet/triplet) can have the same KS density and kinetic energy,
but different KS-pair densities and Hx energies;
(3) KS density and kinetic terms may be expressed as
$n_{s,\FE}=\ibraketop{\FE^{0+}}{\nh}{\FE^{0+}}
=\sum_{i}\theta_{i}^{\FE}|\phi_i|^2$
and $T_{s,\FE}=\sum_{i}\theta_{i}^{\FE}t_i$,
where $\theta_{i}^{\FE}\in\{0,1,2\}$ are occupation factors for
spin-orbital $i$. By contrast,
Hartree-exchange terms $\Lambda_{\Hx,\FE}[\{\phi_i\}]
=\frac12\int d\vr d\vrp W(\vr,\vrp)\allowbreak n_{2\Hx,\FE}(\vr,\vrp)$
must be expressed via the KS-pair densities
$n_{2\Hx,\FE}(\vr,\vrp)=\ibraketop{\FE^{0+}}{\nh(\vr)\nh(\vrp)
- \nh(\vr) \delta(\vr-\vrp) }{\FE^{0+}}$.

Apart from the stated restrictions, so far no approximations
have been made. Thus, we can complete the picture by defining the
correlation energy functional
\begin{align}
  \Ec[n]:=&\F[n]-\F^{\EXX}[n]\;,
  \label{eqn:Ec}
\end{align}
as the difference between the unknown $\F$ and the
exact exchange (EXX) functional $\F^{\EXX}\equiv\Ts+\EHx$.
While formally correct, the above expression has limited
effectiveness in practice. In what follows, we shall introduce
what we argue is a more useful expression for
$\Ec[n]$ {, due to its ability to distinguish
pure-state correlations from those introduced by
ensembles.}

Moving toward this objective, it is important to note
that the KS densities $n_{s,\FE}$ are not the same as the densities
of interacting states $n_{\FE}$. As an example,
consider the lowest lying triplet (ts) and singlet (ss) excited
states in H$_2$. The KS densities of the singlet and triplet
excitation are equal to each other while the interacting ones are
not, i.e. $n_{s,\ts}=n_{s,\ss} {=|\phi_0|^2+|\phi_1|^2}$
{(note, spatial orbitals are the same for spin either up or down)}
and $n_{\ts}\neq n_{\ss}$
\cite{Gould2018-CT}. The same overall ensemble density is,
by construction, obtained from the KS and the real ensemble.
This fact is not
specific to H$_2$, and its implications for the {correlation} energy of ensembles forms the bulk of the remainder of
this letter. We shall first proceed formally, and then review and test
key results in concrete cases.

{\em State- and density-driven ensemble correlations:}
First, it is useful to recall that the energy components can be
restated from functionals of $n$ into functionals of the (ensemble) KS
potential.  As mentioned above, $\Lambda_{\Hx,\FE}$ depends on the
same set of single-particle orbitals as $T_{s,\FE}$ and
$n_{s,\FE}$. Thus, they can all be transformed into a functional of a
potential, by replacing $\phi_i[n]$ by
$\psi_i[v_s] \equiv \phi_i[n[v_s]]$,
where $[\th + v_s]\psi_i[v_s]=\varepsilon_i[v_s]\psi_i[v_s]$.
{Therefore, any functional of the single-particle orbitals
can be readily expressed as a functional of the KS potential;
e.g., $n_{s,\FE}[v_s]\equiv\sum_i\theta_i^{\FE}|\psi_i[v_s]|^2$,
$T_{s,\FE}[v_s]$ and $\Lambda_{\Hx,\FE}[v_s]$.}

As a second and crucial step, we seek to generalize the KS procedure
by finding, for each state $\iket{\FE}$, a unique and state-dependent
KS-like system with effective potential
$\vsE$ such that $n_{s,\FE}[v_s \rightarrow \vsE]=n_{\FE}$ is the
resulting density -- note, $n_{s,\FE}=\sum_i\theta_i^{\FE}|\psi_i[v_s]|^2$
and $n_{\FE}=\sum_i\theta_i^{\FE}|\psi_i[\vsE]|^2$
use the same set of occupation factors.
Finding the corresponding effective potential relies on two
conditions being satisfied: (i) that at least one $\vsE$ exists;
(ii) that multiple valid potentials
(i.e., $v_{s,1}^{\FE},v_{s,2}^{\FE}\to n_\FE$) can
be distinguished through a bi-functional
$\vsE[n_{\FE},n]\equiv\argmin_{\vsE \to n_{\FE}} \|v_s[n],\vsE\|_n$
that selects $\vsE$ as the potential yielding $n_{\FE}$ that is closest to
the true KS potential $v_s$ yielding $n$, according to some measure
$\|v_1,v_2\|_n$ that can depend explicitly on $n$ -- one example is:
$\| v_1,v_2\|_n = \int n(\vr) | v_1(\vr) - v_2(\vr) | d\vr$.

Regarding (i), the two-electron states
considered here (see later discussion) can be mapped to KS
ground-states with well-defined and unique potentials.
KS-like equations for specific eigenstates have
also been introduced to retrieve excitations of Coulomb
systems~\cite{LevyNagy1999,Ayers2015}. Additional
details and discussion appears in the supplementary material.
Regarding (ii), more than one metric may work for the purpose.
This implies some arbitrariness for intermediate quantities
[eqs~\eqref{eqn:EcSD} and \eqref{eqn:EcDD}, below],
yet no difference for their sum [eq.~\eqref{eqn:EcE}].

Once $\vsE$ is determined, we introduce
{$\Tb_{s,\FE}[n_{\FE},n]\equiv  T_{s,\FE}[v_s \rightarrow \vsE[n_{\FE},n]]$
and $\Lambdab_{\Hx,\FE}[n_{\FE},n]\equiv
\Lambda_{Hx,\FE}[v_s \rightarrow \vsE[n_{\FE},n]]$,}
where the original functionals are transformed by replacing the KS
orbitals $\psi_i[v_s]\to\psi_i[\vsE]$ in the
orbital functionals, to give energy bifunctionals of the specific
density $n_{\FE}$ and the total ensemble density $n$.
We thus extend all key functionals to be specified for
ensemble density components, as well as globally.  For the special
case $n_{\FE}=n_{s,\FE}$ we are guaranteed to find
$\vsE[n_{s,\FE},n]=v_s$ by construction. 
It then follows that
$\Ts[n]=\sum_{\FE}w_{\FE}\Tb_{s,\FE}[n_{s,\FE},n]$,
$\EHx[n]=\sum_{\FE}w_{\FE}\Lambdab_{\Hx,\FE}[n_{s,\FE},n]$.

Finally, {we can express the correlation energy as:
\begin{align}
  \Ec[n] =& \Ec^{\SD}[n] + \Ec^{\DD}[n],
  \label{eqn:EcE}
\intertext{where}
  \Ec^{\SD/\DD}[n] \equiv & \sum_{\FE}w_{\FE}\Eb_{\crm,\FE}^{\SD/\DD}[n_{\FE},n]. 
  \label{eqn:EcEb}
\end{align}%
}%
Here, the {``pure''} state-driven (SD),
\begin{align}
  \Eb_{\crm,\FE}^{\SD}[n_{\FE}, n] :=& \bar{F}_{\FE}[n_{\FE},n] -
  \FEXXb[n_{\FE},n],
  \label{eqn:EcSD}
  \intertext{and {``ensemble''} density-driven (DD),}
  \Eb_{\crm,\FE}^{\DD}[n_{\FE},n] :=& \FEXXb[n_{\FE},n] -
  \FEXX[n]
  \label{eqn:EcDD}
\end{align}
terms are defined {using}
$\bar{F}_{\FE}[n_\FE,n] := E_{\FE}[n]-\int d\vr n_{\FE}(\vr)v[n](\vr)$,
$\FEXXb[n_{\FE},n] := \Tb_{s,\FE}[n_{\FE},n] +
\Lambdab_{\Hx,\FE}[n_{\FE},n]$, and  $\FEXX[n] := T_{s,\FE}[n] +
\Lambda_{\Hx,\FE}[n] \equiv \FEXXb[n_{s,\FE},n]$
(since $n_{s,\FE}$ depend on $v_s[n]$).

Eq.~\eqref{eqn:EcE} is the key result of the present work. It
expresses the correlation energy of GOK ensembles in terms of:
(a) state-driven correlations {[eq. \eqref{eqn:EcSD}]}
which are like the usual pure state correlation energy,
but involve bifunctionals of $[n_{\FE},n]$;
{\em and} (b) density-driven correlations
{[eq. \eqref{eqn:EcDD}]},
which resemble difference between exact exchange energies at
different pure state densities.
{The labelling of SD terms as ``pure'' and DD as ``ensemble''
can now be explained. In a pure state, $n_{s,\gs}=n_{\gs}=n$ and
thus $\Ec^{\DD}=0$, as expected. Moreover,
in \emph{any} ensemble, the ground-state term
$\bar{E}_{c,\gs}^{\SD}$ depends only on $n_{\gs}$, and not on $n$
(since $v_s^{\gs}$ is unique). By contrast,
$\bar{E}_{c,\gs}^{\DD}$ always depends on both $n$ and $n_{\gs}$,
so varies with the overall choice of ensemble. Density-driven
correlations are consequently a unique, yet unavoidable, feature
of EDFT -- they appear because the KS system cannot
simultaneously reproduce the densities of all ensemble components.}

{\em Implications:}
First of all, our decomposition
{need not handle problematic self- or ghost-} interactions
\cite{Pastorczak2014-GI,Pribram-Jones2014-GI,Gidopoulos2002-GI}.
Because, our correlation functional is defined on top of
an ensemble Hartree-exchange which is already maximally free from
such {spurious} interactions.
Any spurious interactions present must thus
be the result of approximation. Our decomposition, of course,
is not meant to tame unavoidable strong correlations
{in the SD terms}.

We now turn to how our scheme can help
in the development of new approximations. Inspired by the
principle of minimal effort, one might seek to replace
the entire correlation energy with the SD terms, eq.~\eqref{eqn:EcSD},
by reusing any standard DFT approximation (DFA),
i.e. set $E_{\crm,\FE}^{\SD}[n_{\FE},n] \rightarrow
E_{\crm}^{\rm DFA}[n_{s,\FE}]$. The idea of reusing standard DFAs in
ensembles is not new in EDFT,
and with appropriate care has
been shown to give good results in excited state and related
non-integer ensembles
\cite{Kraisler2013,Pastorczak2014-GI,Filatov2015-Double}.
In the present context [see eq.~(\ref{eqn:EcE})
{and eq.~(\ref{eqn:EcEb})}], however,
we can appreciate that such a procedure:
(a) replaces the interacting densities of the SD terms by their
non-interacting counterparts, to make use of ingredients that are
available in a typical calculations;
(b) disregards the additional functional dependence of the SD terms on
$n$; and (c) misses the DD terms entirely.

Next, we show that the contribution of the DD terms are indeed of
relevant magnitude, when all the exact quantities are evaluated
numerically. Then, we shall {discuss approximations}.

\begin{figure}[b!th]
  \includegraphics[width=0.9\linewidth]{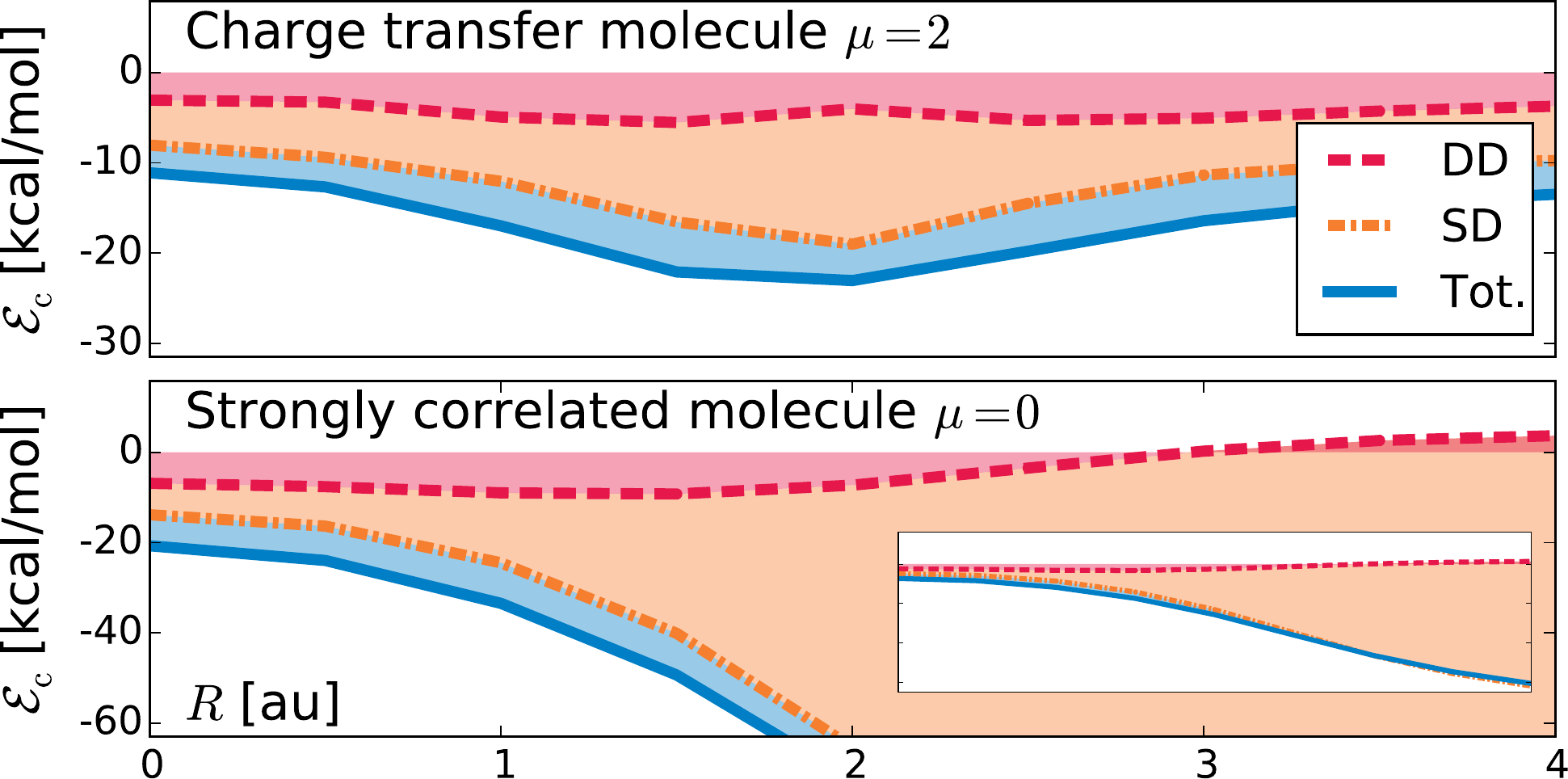}
  \caption{Decomposition of the correlation energy of the
    charge transfer (top) and strongly-correlated (bottom)
    cases. The shaded regions show the relative significance of
    density-driven and state-driven correlations, with the
    former contributing approximately one quarter of the total
    correlation energy in the charge transfer case.
    The inset of the bottom panel {illustrates} the unzoomed plot.
    Here we set a mixture of 60/30/10\% respectively for the
    three lowest energy states.
    \label{fig:EcDDens}}
\end{figure}

{\em Applications:}
Having established the basic theory, let us now study the role of
density-driven correlations in two electron
soft-Coulomb molecules. {These tunable (via parameter $\mu$)
one-dimensional molecules can exhibit
chemically interesting properties such as charge transfer
excitations ($\mu=2$) or strong correlations ($\mu=0$)
\cite{Gould2018-CT} and thus allow important physics to
be analyzed with full control.
Details are in the in the Supplementary Material.}

We restrict ourselves to ensembles involving the ground- (gs),
triplet-excited (ts) and singlet-excited (ss) states only.
We perform our calculations in three steps:
\\{\bf Step 1:}
Solve the two electron Hamiltonian $\Hh$ with one- and
  two-body interactions terms to obtain interacting state-specific
  terms $E_{\FE}$, $\iket{\FE}$, $n_{\FE}$,
  $F^1_{\FE}=\ibraketop{\FE}{\Th+\Wh}{\FE}=E_{\FE}-\int dx
  n_{\FE}(x)v(x)$, for the three states $\FE\in\{\gs,\ts,\ss\}$, and
  ensemble averages therefrom, e.g., $n=\sum_{\FE}w_{\FE}n_{\FE}$ and
  $\F^1=\sum_{\FE}w_{\FE}F^1_{\FE}$.
\\{\bf Step 2:}
  Invert\cite{Gould2014-KS} the density using the single-particle
  orbital Hamiltonian $\hh=-\half\partial_x^2 + v(x)$ to find
  $v(x)=v_s(x)\to n(x)$ and real-valued
  orbitals $\phi_0$ and $\phi_1$ that
 are required for the KS eigenstates. Here, $v_s$ depends on the
  density $n$ and groundstate weight $w_{\gs}$ only,
  as $n=(1+w_{\gs})\phi_0^2 + (1-w_{\gs})\phi_1^2$.
  From these terms, calculate $n_{s,\FE}$, $n_{2\Hx,\FE}$
  $T_{s,\FE}$ and $\Lambda_{\Hx,\FE}$, and ensemble averages, again
  for $\FE\in\{\gs,\ts,\ss\}$. Here, $T_{s,\ts}=T_{s,\ss}$ and
  $n_{s,\ts}=n_{s,\ss}$ but $\Lambda_{\Hx,\ts}\neq\Lambda_{\Hx,\ss}$
  and $n_{2\Hx,\ts}\neq n_{2\Hx,\ss}$.
\\{\bf Step 3:}
  Carry out separate inversions using
  $n_{\gs}=2\psi_0[v_s^{\gs}]^2$,
  $n_{\ts}=\psi_0[v_s^{\ts}]^2+\psi_1[v_s^{\ts}]^2$ and
  $n_{\ss}=\psi_0[v_s^{\ss}]^2+\psi_1[v_s^{\ss}]^2\neq n_{\ts}$ to
  obtain the three unique potentials $\vsE$.
  Then use the resulting orbitals $\psi_0[\vsE]$ and
  $\psi_1[\vsE]$ to calculate $\Tb_{s,\FE}[n_{\FE},n]$ and
  $\Lambdab_{\Hx,\FE}[n_{\FE},n]$ on the interacting densities of
  the three states, and thus obtain the final ingredients for
  eqs~\eqref{eqn:EcE}--\eqref{eqn:EcDD}.

In Figure~\ref{fig:EcDDens} we show the correlation energy for two
examples of bond breaking (which occurs at $R\approx 3$),
resolved into total, DD and SD components.
One example exhibits charge transfer excitations (top, $\mu=2$),
and the other involves strong correlations (bottom, $\mu=0$).
We choose an ensemble with 60\% groundstate,
30\% triplet state and 10\% singlet state (60/30/10\%).

The first thing to notice is that in the ``typical'' charge
transfer case, the DD correlations form a substantial portion of the
total correlation energy, about 25\% on average. This highlights the
importance of capturing, or approximating it somehow: a raw
application of even a nearly perfect approximation to the SD
correlations will miss around one quarter of the correlation energy.
The strongly correlated case has a similar breakdown for small $R$, but
becomes dominated by the SD correlations for large $R$.
This is not surprising, as
the SD term captures the multi-reference physics that gives rise
to most of the correlation energy, whereas the DD term contains
only weaker dynamic correlations.
The various densities that give rise to the DD correlations
are shown and discussed in the Supplementary Material.

Of final note, close inspection of the strongly correlated case reveals
a subtle point: for $R\geq 3$, the DD correlation
energy is \emph{positive}. At first glance this might seem
to be impossible -- correlation energies should always be
negative. However, it reflects the fact that the DD
correlation energy is defined via an energy difference between two
states which come from different many-body problems with different
densities. Thus, the negative sign is not guaranteed by any
minimization principle.

\begin{figure}
  \includegraphics[width=0.9\linewidth]{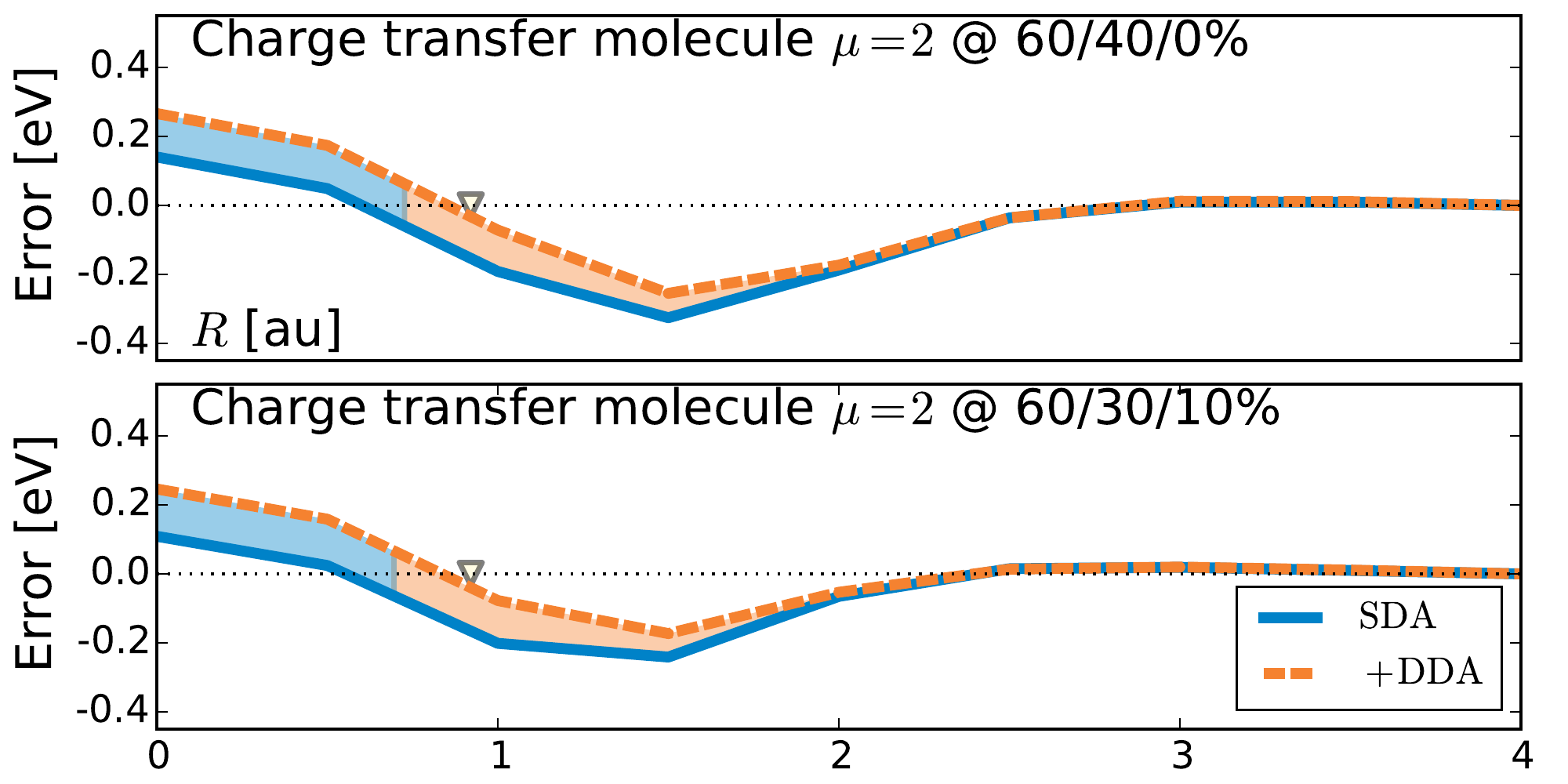}
  \caption{The error
    $\text{Err}(R) = \Delta\EHxc^{\rm approx}-\Delta\EHxc^{\rm exact}$
    in excitation energies $\Delta\E = \E - E_{\gs}$, shown relative
    to the dissociation limit,
    $\text{Err}(R=4)$. Shown are 60/40/0\% (top) and
    60/30/10\% (bottom) mixtures for the charge transfer case.
    We report both the pure state-driven only approximation (SDA) and 
    the SD term plus the DD approximation (+DDA).
    Shaded regions indicate where including the DD approximation
    improves on the SD-only case (orange), or worsens it (blue).
    The diamond indicates the equilibrium interatomic distance.
    \label{fig:Approx} }
\end{figure}

So far we have been concerned with exact quantities. But for
applications, it is essential to derive approximations.
For a proof-of-principle demonstration, let us  focus on
charge transfers in 1D molecules.
{We approximate the SD terms using available ingredients 
for our 1D model -- working in 3D would let us
generate a variety of forms by tapping into the existing
DFT zoo. The reported approximations use numerically
exact KS densities $n_{s,\FE}$.}

We generate a SDA by  combining the ensemble exact Hx results with
a local spin density approximation (LSDA) for correlation,
parametrised for the 1D soft-Coulomb potential
\cite{Helbig2011,Wagner2012,Casula2017-PC}.
But we adapt the LSDA  according to the formalism
laid out by Becke, Savin and Stoll \cite{Becke1995} -- which is useful
for dealing with multiplets. Full details are provided in the
Supplementary material.

The key point to be addressed here is the
approximation for the DD terms (DDA). As far as charge transfer are
concerned, intuition suggests that an electrostatic model may
work well for a first DDA. Thus, we propose
$\Ec^{\DDA}= \sum_{\FE}w_{\FE}
\big\{ E_{\Hrm}[ n_{\FE} \rightarrow \nt_{\FE}] - E_{\Hrm}[n_{s,\FE}] \big\}$.
This expression involves the KS densities $n_{s,\FE}$ and 
$\nt_{\FE}= S_{\FE} n_{s,\FE}[1
  + a\Delta n_{s,\FE} + b\Delta n_{s,\FE}^2]$
which accounts for the
fact that in real situations we may not access the exact $n_{\FE}$.
Here, $S_{\FE}$ is chosen to ensure the correct number of electrons,
and the term $\Delta n_{s,\FE}=n_{s,\FE}-n$ (i.e., the deviation of
the state density $n_{s,\FE}$ from the full ensemble density $n$),
ensures that the correction is zero in the case of a pure state.
Parameters $a=-0.28$ and $b=0.12$ are found via optimization.
Additional information on our DDA, including comparisons
with the exact DD term, are provided in the Supplementary
material.

Figure~\ref{fig:Approx} shows errors in our approximations for
the 60/30/10\% case from earlier, and a 60/40/0\% case without
singlet excitations. Although the proposed approximation
neglects both kinetic and x-like contributions [see
  eq.~(\ref{eqn:EcDD})], its performance is remarkably good.
Including the DDA improves results for almost all chemically
relevant $R$ (see orange shading).

{{\em Summary and outlook}: Correlations in ensemble density
  functional theory (EDFT) are more than the simple sum of their
  parts. They naturally divide
  into state-driven (SD) and
  density-driven (DD) contributions, the former being amenable to
  direct translation of existing DFT approximations, and the latter
  being a unique property of ensembles.  In prototypical ensembles of
  excited states, DD correlations account for up to 30\% of the
  overall correlation energy. Therefore, accurate approximation
  of the correlation energy requires simultaneous
  consideration of the SD and DD components.}

{A simple approximation
  to the DD correlations was devised and evaluated in model
  situations. Thus, accounting for both
  SD and DD correlations was
  shown to be both feasible and promising to prompt progress in EDFT.
  Development of general approximations, extension to deal with
  systems that may challenge our simplifying ``strong adiabatic''
  assumption, and generalization of key concepts and procedures
  presented here to other ensembles
  \cite{Perdew1982,Gould2013-LEXX,Senjean2018,Deur2019} are being
  pursued.}

\bibliography{DDCorrelations}

\end{document}


\title{Supplementary Material for \\ ``Density-driven correlations in many-electron ensembles: \\ theory and application for excited states''}
\author{Tim Gould}\affiliation{Qld Micro- and Nanotechnology Centre, %
  Griffith University, Nathan, Qld 4111, Australia}
\author{Stefano Pittalis}\affiliation{CNR-Istituto Nanoscienze, Via
  Campi 213A, I-41125 Modena, Italy}

\begin{abstract}
We provide further information on: (i) auxiliary (KS-like)
non-interacting systems for excited pure-state interacting densities;
{(ii) The 1D model system used in numerical examples;}
(iii) KS and exact densities for pure states in ensembles;
(iv) the approximations introduced in the manuscript.
\end{abstract}

\maketitle

\section{On the auxiliary non-interacting systems for excited pure-state interacting densities}

In the paper, we briefly discuss circumstances in which effective
KS-like potentials can be shown to exist for excited pure-state
interacting densities. Here, we expand on this
discussion. Since most systems of physical or chemical interest
exhibit ``nice" particle densities, we shall not go into the
discussion of peculiar or subtle cases. 
Note, this `machinery' is introduced in our work to decompose the
energy correlation into the  state-driven and density-driven energy
correlations. The orbitals obtained from the aforementioned
KS-like potentials are required to calculate
$\bar{F}^{\EXX}[n_{\FE},n]$. Other terms in the decomposition can be
either regarded  as functionals of the densities alone or can be
computed by using standard (ensemble) KS orbitals. All the required
definitions are given within the paragraph stating equation~(8) in
the main text.

Firstly, in the case of  excited states of \emph{Coulomb systems}
we can use the work in Ref.~\onlinecite{LevyNagy1999,Ayers2015}
to provide a general proof.
We can follow the arguments of
\rcite{Ayers2015}, to obtain $T_{s,\FE}[n_{\FE}]$ from their
equation~(8). Then, since, equation~(12) provides the potential
$\vsE$ for $n_{\FE}$ we can obtain the orbitals
$\psi_i[\vsE]$ for use in
$\bar{F}^{\EXX}_{\FE}[n_{\FE},n]=T_{s,\FE}[n_{\FE}]
+ \Lambda_{\Hx,\FE}[\{\psi_i[\vsE]\}]$.

Secondly, the cases in the manuscript are for \emph{soft-Coulomb systems} and are not amenable to the above treatment.
However, they involves an ensemble composed only of a doubly-occupied
singlet ground-state, and the first-lying triplet and singlet states
formed by promotion of a single orbital. For these states, we show
here, that the particle densities may be retrieved by using
single-particle orbitals of non-interacting \emph{ground-state}
problems -- in the sense to be specified below.

The case of the actual ground states are obvious and, thus, do not
need to be discussed further. For the lowest-lying triplets, we only
need to observe that they behave effectively as ground states in the
minimizations of the energy within the given (triplet)
multiplicity. This is, in fact, an old trick used in the KS literature
to access lowest-lying states of any prescribed multiplicity.

This trick, however, does not apply to the first-lying
excited singlets (in our examples, the actual ground states are
singlets). But at the level of \emph{non-interacting} states, 
singlet and triplet lead to particle densities of the same form
[see, for example, eq.~\eqref{eqn:ns} below]. 
Thus, we may retrieve the
particle density as well as the  single-particle orbitals for
our excited \emph{interacting} singlet (which differs from the
particle density of our excited \emph{interacting} triplet) from the
lowest-lying triplet of a \emph{different} system.
Hence, the corresponding single-particle
orbitals for all three states may be determined uniquely by
usual KS inversion routines\cite{Gould2014-KS}. The orbitals thus
obtained can then be used to calculate $\bar{F}^{\EXX}[n_{\FE},n]$.

Finally, there is yet another way to rely on standard KS inversion
procedures. Our specific case is for \emph{two} interacting
electrons. For it, we may search for the auxiliary local potential
that has a closed-shell ground state for \emph{four}
\emph{non}-interacting electrons having double the prescribed
particle density. This state has same orbitals as the desired
two-electron \emph{non-interacting} excited triplet/singlet but with
doubled occupations [to see this mathematically, double the
second expression in eq.~\eqref{eqn:ns}, below, to get
$n^{\text{4-el}}\equiv 2n_{\ts/\ss}=2|\phi_0|^2 + 2|\phi_1|^2$,
which is the usual four-electron KS density expression].
Thus, we can eventually obtain
$\bar{F}^{\EXX}[n_{\FE},n]$.

Therefore, the existence (and, indeed, uniqueness) of the
potentials is guaranteed in most cases of interest.
We also note, in passing,  that the ability of the various Coulomb
functionals of \rcites{LevyNagy1999,Ayers2015} to deal with
\emph{specific} excited states might also offer a route for
bypassing the strong adiabatic assumption
which is so far  required in our work.

{\section{1D Soft-Coulomb molecules}}

{
The 1D soft-Coulomb molecules we study, both in the manuscript
and here, involve a Hamiltonian
\begin{align}
  \Hh=-\half[\partial_{x_1}^2+\partial_{x_2}^2]
  +W(x_1-x_2)+v(x_1)+v(x_2)\;,
\end{align}
in one spatial dimension. Here, the
electron-electron interaction term is
$W(x)=(\frac14+x^2)^{-\half}$. The parametrised external
potential is $v(x)=-W(x+R/2)-W(x-R/2)-\mu e^{-(x-R/2)^2}$,
for nuclear distance $R$ and adjustable well-depth $\mu$
on the right atom.}

{This model is reasonably straightforward to solve numerically.
More importantly, by varying the parameter $\mu$ (the effective
well-depth on the right atom) we can {explore its behaviour}
from strongly correlated (using $\mu=0$) to charge transfer
(using $\mu=2$) states.}

\section{KS and exact densities for pure states in ensembles}

In the manuscript we highlight the effect of density-driven
correlations on energies, both in exact and approximate cases. Here we
analyze the various particle densities for the two cases shown in
Figure~1 of the manuscript.

Note, in this section `KS' strictly refers to its original meaning as
intended in EDFT. Therefore, the KS system which yields a prescribed
ensemble particle density does not need to reproduce the particle
densities of each pure state in the same ensemble.
The two-electron systems we consider
have {\em degenerate} (for the triplet/singlet) 
KS densities of the form
\begin{align}
  n_{s,\FE}(x)=&\begin{cases}
    2|\phi_0(x)|^2, & \FE=\text{gs}, \\
    |\phi_0(x)|^2+|\phi_1(x)|^2, & \FE=\text{ts/ss}.
  \end{cases}
  \label{eqn:ns}
  \end{align}
where $\phi_0(x)$ and $\phi_1(x)$ are the required single-particle
orbitals.
  
\begin{figure*}[htb]
  \includegraphics[width=\linewidth]{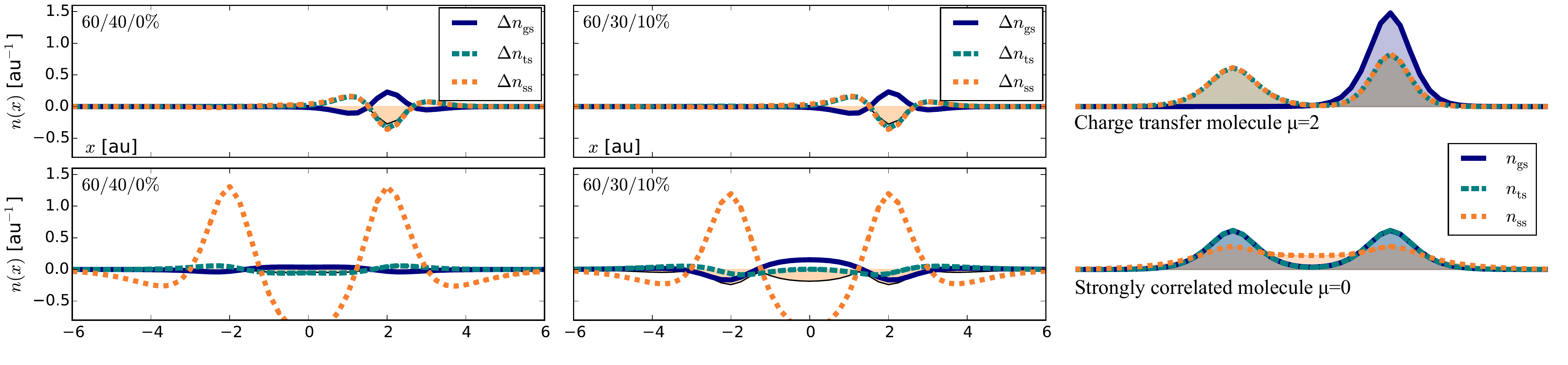}
  \caption{Density differences for charge transfer (top)
    and strongly-correlated (bottom) cases, and with only triplet
    states (left) or with singlet states as well (centre). The
    line plots show $5\times$ the density difference for each
    state (navy, teal and orange dashed lines), and the cream
    shaded area shows $-5\times$ the weighted average absolute
    density difference $\Delta|n|$. The final panel
    (right) shows the true densities of the three states,
    for visual comparison.
    \label{fig:DDens}}
\end{figure*}

Thus, Supplementary~Figure~\ref{fig:DDens} shows density
differences $\Delta n_{\FE}=n_{s,\FE}-n_{\FE}$ between KS and true
states in the left and 
middle panels, and the true densities at the right. It also shows the
weighted mean absolute density difference $\Delta|n|
=\sum_{\FE}w_{\FE}|\Delta n_{\FE}|$, to visually summarise the density
difference that may affect the energy, keeping in mind that
$\sum_{\FE}w_{\FE}\Delta n_{\FE}=0$. In all cases it is clear that the
density differences are substantial.

Specifically, Supplementary Figure~\ref{fig:DDens} shows
results for the case $R=4$ with $\mu=0$ (strong correlations)
and $\mu=2$ (charge transfer). It
 includes the 60/30/10\% case (middle) analysed in the main
manuscript, but also a 60/40/0\% case (left) without any singlet
contribution. The singlet-free case lets us explore a subtle point on
how the weights affect the densities.

One particularly interesting feature is that the strongly
correlated case (bottom) shows fundamentally different deviations when
the singlet is neglected or included, reflecting the large errors
$n_{s,\ts}-n_{\ts}$ and $n_{s,\ss}-n_{\ss}$ in simultaneously trying to
represent the triplet and singlet states using
$n_{s,\ts}=n_{s,\ss}=|\phi_0|^2+|\phi_1|^2$.
Any calculation of ensembles involving the three lowest
energy configurations will need to handle such a difficult case
via a direct density-driven correlation energy approximation.

\section{Definition of approximate correlation forms}

\subsection{State-driven contributions}

We set out to approximate the state-driven correlations using the
exact ensemble Hartree-exchange (HX) plus a modification of the
correlation energy as defined in local spin density approximation
(LSDA) [see eq.~(\ref{eq:SDA}) below]. Since we have discuss the Hx
expression in detail in the manuscript, let us first focus on
correlation.

The correlation energy per electron in the LSDA for 1D soft-Coulomb
systems takes the form \cite{Helbig2011,Wagner2012}
\begin{align}
  \epsilon_{\crm}(r_s)\approx& -\half
  \frac{r_s+Er_s^2}{A+Br_s+Cr_s^2+Dr_s^3}\log(1+\alpha r_s+\beta r_s^m)\;.
  \label{eqn:epsc}
\end{align}
where $r_s=1/(2n)$. Variations due non-vanishing polarization,  $\zeta=(n_{\up}-n_{\down})/n$, are accounted for with
\begin{align}
  \epsilon_{\crm}(r_s,\zeta)=&(1-\zeta^2)\epsilon_{\crm}(r_s,\zeta=0)
  + \zeta^2\epsilon_{\crm}(r_s,\zeta=1).
\end{align}
The results reported in \cite{Helbig2011} and \cite{Wagner2012} are
for $W'=(1+X^2)^{-\half}$, whereas we use $W=(\frac14 +X^2)^{-\half}$.
Unpublished parameters for $\epsilon_{\crm}$ for our case  were
obtained by private communication\cite{Casula2017-PC}: $A=7.4070$,
$B=1.11663$, $C=1.8923$, $D=0.0960119$, $E=0.024884$,
$\alpha=2.43332$, $\beta=0.0142507$ and $m=2.9198$
(for $\zeta=0$);
and $A_1=5.248$
$B_1=0$, $C_1=1.568$, $D_1=0.1286$, $E_1=0.00321$,
$\alpha_1=0.0539$, $\beta_1=0.0000156$, and $m_1=2.959$
(for $\zeta=1$).

We can adapt the LSDA such to use $\zeta$ to distinguish the various
states  yet preserving their multiplet structure.
This may be implemented with the replacement
\begin{align}
 \zeta(x) \rightarrow~  \zeta_{\FE}(x)=&\sqrt{ \max\big[0,
      1 - \frac{2n_{2\Hx,\FE}(x,x)}{n_{\FE}(x)n_{\FE}(x)}\big] }\;,
  \label{eqn:zeta}
\end{align}
where $n_{2\Hx,\FE}(x,x)$ stands for the pair density of
non-interacting KS states (see in the main manuscript).  In doing
this, we are borrowing aides from the work of Becke, Savin and Stoll
\cite{Becke1995} and make the additional request that imaginary values
are avoided by means of the $max$-function -- note, this expression
provides the exact polarization for single Slater determinants.

Let us see in detail how we deal with the states analyzed in the
previous section. Readily, we find
\begin{align}
  n_{2\Hx,\FE}(x,x')=&\begin{cases}
    2|\phi_0(x)|^2|\phi_0(x')|^2, & \FE=\text{gs}, \\
    |\phi_0(x)\phi_1(x')-\phi_1(x)\phi_0(x')|^2,
    & \FE=\text{ts}, \\
    |\phi_0(x)\phi_1(x')+\phi_1(x)\phi_0(x')|^2,
    & \FE=\text{ss}.
  \end{cases}
  \label{eqn:n2Hx}
 \end{align}
Through eqs~\eqref{eqn:ns}, \eqref{eqn:zeta} and \eqref{eqn:n2Hx},
we obtain $\zeta_{\gs}= 0$, $\zeta_{\ts} = 1$, and
$\zeta_{\ss}=\sqrt{\max[0,(n_0 - n_1)^2 -4n_0n_1]}/(n_0+n_1)$.

Hence, our state-driven approximation is defined by the expression 
\begin{align}\label{eq:SDA}
  \EHxc^{\SDA}=&\sum_{\FE}w_{\FE} \left\{
\Lambda_{\Hx,\FE}
  + \int dx n_{s,\FE}(x) \epsilon_{\crm}(r_{s,\FE}(x),\zeta_{\FE}(x)) \right\},
\end{align}
where $\Lambda_{\Hx,\FE}$ is given in the main manuscript.
Because eq.~(\ref{eq:SDA}) is the only state-driven approximation we use in the main text, no risk of confusion may  emerge
by  denoting its expression compactly and simply as SDA.

\subsection{Density-driven contributions}

In introducing a first approximation for the density-driven
correlations (DDA), we  allow ourselves a ``minimalistic''  approach.
As discussed in the manuscript, a density driven term essentially
involves the difference between
$\bar{T}_s+\bar{E}_{\Hx}$ calculated at interacting density $n_{\FE}$ and
Kohn-Sham density $n_{s,\FE}$. It thus involves a complicated orbital
dependence. To avoid this, we assume for simplicity that the
difference in the kinetic energy and exchange contributions may be
small, relative to the simpler electrostatic term, so that
\begin{align}\label{eq:DDA-H}
  \Ec^{\DDA} =& \sum_{\FE}w_{\FE}\big\{ E_{\Hrm}[n_{\FE}]
  - E_{\Hrm}[n_{s,\FE}]  \big\}.
\end{align}
But in a typical calculation we would
not have access to the exact $n_{\FE}$. Thus we replace it 
with
\begin{align}
  n_{\FE} \rightarrow~ \nt_{\FE} = S_{\FE} n_{s,\FE}
  \big[1 + a\Delta n_{s,\FE} + b\Delta n_{s,\FE}^2\big].
  \label{eqn:ntFE}
\end{align}
The term $S_{\FE} = N_e/(N_e+ aM_{\FE,1} + bM_{\FE,2})$, where 
$N_e=\int n(x)dx$, $\Delta n_{s,\FE}=n_{s,\FE}-n$, and
$M_{\FE,p} = \int n_{s,\FE}(x) \Delta n_{s,\FE}^p(x) dx$,
is chosen to ensure the correct number of electrons.
$\Delta n_{s,\FE}=n_{s,\FE}-n$ is employed to
guarantee no contribution in the case of a pure state, where it must
be zero since $n_{\FE=0}=n_{s,\FE=0}=n$. Note, our goal is not to find
accurate densities, but to find an accurate approximation for the
density-driven correlation energy $\Ec^{\DD}$.

Finally, the two parameters $a=-0.28$ and $b=0.12$ are found by
minimizing the error
$\min_C\sum_R|\Ec^{\DDA}(R)-\Ec^{\DD}(R)-C|$ for all $R$
over 10 dissociation curves (90 calculations in total for 9 values
of $R$ in each). Here, we allow for a systematic deviation
(via constant $C$) with the aim of yielding a good approximation
for the more-important energy differences,
e.g. the approximate dissociation curves.
The benchmarks cover all $R\in(0,1/2,1,\ldots 4)$, $\mu\in (0,2)$,
and $\W\in(90/10/0\%, 80/20/0\%, 70/30/0\%, 60/40/0\%,
60/30/10\%)$ -- we thus seek to reduce the risk of overfitting for our
test cases.

Here Supplementary Figure~\ref{fig:EcDD} shows results from the DDA
and from the exact calculations, shown relative to their values in the
dissociation limit, i.e. $R=4$ for all practical purposes. 
The approximation is almost perfect in the 60/40/0\% case, but less
good in the 60/30/10\% case, see discussion below. Nonetheless, it
gives values within 0.2~eV of the exact results even in its worst
cases.

The 60/30/10\% case identifies a major limitation of our approximation: 
the density-driven correction for the singlet and
triplet excitations are necessarily the same, as their KS densities
are the same. This means the approximation gives essentially the same
results for the 60/30/10\% and 60/40/0\% cases, despite the two having
different exact $\Ec^{\DD}$ (the small difference is caused by the
different state and overall densities). Future approximations might
use the pair-density $n_{2\Hx,\FE}$ of the states or possibly
$\zeta_{\FE}$, as above in \eqref{eqn:zeta}, to improve their
flexibility.

\begin{figure}[b!th]
  \includegraphics[width=\linewidth]{{{GGADensDriven_CTDDAError}}}
  \caption{Deviations $\Ec(R)-\Ec(R=4)$ of the density-driven
    approximation $\Ec^{\DDA}$ compared to the exact results
    $\Ec^{\DD}$. Also shown are the errors
    $\Delta\Ec^{\DDA}=\Ec^{\DDA}-\Ec^{\DD}$ of the approximation,
    which are less than 0.1~eV in all cases.
    \label{fig:EcDD}}
\end{figure}

\bibliography{DDCorrelations}